\documentclass[aps,prb,reprint,superscriptaddress]{revtex4-1}

\usepackage{graphicx}
\usepackage[version=4]{mhchem}
\usepackage{float} % Required for tables and figures in the multi-column 
\usepackage[normalem]{ulem}
\usepackage[hidelinks]{hyperref} % For hyperlinks in the PDF
\usepackage{paralist}
\usepackage{dcolumn} % Align table columns on decimal point
\usepackage{bm}
\usepackage[usenames,dvipsnames,svgnames,table]{xcolor}
\usepackage[flushleft]{threeparttable}

\renewcommand{\vec}[1]{\mathbf{#1}}

\newcommand{\tauanh}{\tau^{\textrm{anh}}_{j\mathbf{q}}}

\newcommand{\taumd}{\tau^{\textrm{md}}_{j\mathbf{q}}}
\newcommand{\taubs}{\tau^{\textrm{bs}}_{j\mathbf{q}}}

\begin{document}
\title{Lattice thermal conductivity of Ti$_x$Zr$_{y}$Hf$_{1-x-y}$NiSn half-Heusler alloys calculated from first principles:
Key role of nataure of phonon modes }
\author{Simen N. H. Eliassen}
%\affiliation{Centre for Materials Science and Nanotechnology (SMN), Department of Physics, University of Oslo, Norway}
\affiliation{Centre for Materials Science and Nanotechnology, Department of Physics, University of Oslo, Norway}
\affiliation{Department of Materials Science and Engineering, Norwegian University of Science and Technology, Norway}

\author{Ankita Katre}
\affiliation{LITEN, CEA-Grenoble,  France}
\author{Georg K. H. Madsen}
\affiliation{Institute of Materials Chemistry, TU Wien, Austria}

\author{Clas Persson}
\affiliation{Centre for Materials Science and Nanotechnology, Department of Physics, University of Oslo, Norway}

\author{Ole Martin L\o vvik}
\affiliation{Centre for Materials Science and Nanotechnology, Department of Physics, University of Oslo, Norway}
\affiliation{SINTEF Materials and Chemistry, Norway}

\author{Kristian Berland}
\affiliation{Centre for Materials Science and Nanotechnology, Department of Physics, University of Oslo, Norway}

\date{\today}

\begin{abstract}
  In spite of their relatively high lattice thermal conductivity $\kappa_{\ell}$, the \textit{X}NiSn (X=Ti, Zr or Hf) half-Heusler compounds are good thermoelectric materials. 
  Previous studies have shown that $\kappa_{\ell}$ can be reduced by 
  sublattice-alloying on the \textit{X}-site.
To cast light on how the alloy composition affects $\kappa_\ell$, we 
study this system using the phonon Boltzmann-transport equation within the relaxation time approximation
 in conjunction with density functional theory.
The effect of alloying through mass-disorder scattering is explored using the virtual crystal approximation to screen the entire ternary Ti$_x$Zr$_{y}$Hf$_{1-x-y}$NiSn phase diagram. 
The lowest lattice thermal conductivity is found for the Ti$_x$Hf$_{1-x}$NiSn compositions; in particular, there is 
a shallow minimum centered at Ti$_{0.5}$Hf$_{0.5}$NiSn with $\kappa_l$ taking values between 3.2 and 4.1 W/mK when the Ti content varies between 20 and 80{\%}.
Interestingly, the overall behaviour of mass-disorder scattering in this system can only be understood from a combination of the nature of the phonon modes and the magnitude of the mass variance. 
Mass-disorder scattering is not effective at scattering acoustic phonons of low energy.
By using a simple model of grain boundary scattering, we find that nanostructuring these compounds can scatter such phonons effectively and thus further reduce the lattice thermal conductivity; for instance, Ti$_{0.5}$Hf$_{0.5}$NiSn with a grain size of $L= 100$ nm experiences a 42\% reduction of $\kappa_{\ell}$ compared to that of the single crystal. 
\end{abstract}


\maketitle

%------------------------------------------------ %

\section{Introduction}\label{sec:intro}

% Thermoel general
With their ability to harvest waste heat, thermoelectric materials can help reducing the global energy consumption.\cite{Yang2009, Bell1457}
A thermoelectric device typically contains pairs of n- and p-type semiconductors, generating an electric current from a difference in temperature $T$.
The energy conversion efficiency of a thermoelectric material is characterized by the dimensionless figure of merit,
\begin{equation}\label{eq:zt}
  zT = \frac{\sigma S^2}{\kappa_{\textrm{e}} + \kappa_{\ell}}T,
\end{equation}
where $ \sigma $ is the electrical conductivity, $ S $ is the Seebeck 
coefficient, and $ \kappa_{\textrm{e}}$  and $ \kappa_{\ell} $ are the electronic and lattice 
thermal conductivity. 
Thus a high power factor $ \sigma S^2 $ is desirable whereas the thermal conductivity $\kappa$ should be as low as possible.
As $ \kappa_{\textrm{e}} $ is roughly proportional to $ \sigma $ through the Wiedemann-Franz law and often smaller than  $ \kappa_{\ell} $ in semiconductors, reducing the latter is an important strategy. 

% HH general 
Half-Heusler (HH) compounds with the general formula \textit{XYZ}, crystallize in the $F\bar{4}3m$ spacegroup. The \textit{XZ} sublattice forms a rock-salt structure of which half the interstitial sites are occupied by \textit{Y}, so that the \textit{YZ} sublattice forms a zinc-blende structure. 
There are many ways to combine three different elements to form  HH compounds,\cite{carrete2014} even when restricted by thermodynamic considerations.\cite{Zunger2015} 
This flexibility opens for many potential applications, ranging from spintronics\cite{Casper2012_review} to optoelectronics.\cite{Gruhn2010_opto}  
In particular, the many possibilities allow for the use of inexpensive, earth-abundant and environmentally friendly elements.
This is one of the reasons why HH compounds have attracted much attention lately as potential thermoelectric materials. \cite{Esfarjani2011,bos_and_downie2014,ChenRen2013,HHZintl:Zeier,Yang:HH_creening,carrete2014,AENM:AENM201500588,XieWeidenkaffTangEtAl2012}
The $n$-type \textit{X}NiSn compounds are particularly efficient in the temperature range 400--1000~K and thus hold an important niche where few other competing alternatives exist. \cite{ChenRen2013,zhifeng2011,poon2011}
These compounds often exhibit a high Seebeck coefficient and high electrical conductivity, and thus a high power factor, while their main disadvantage is a relatively high thermal conductivity. \cite{shen2001,Uher1999,hohl1999,tritt2001,Kawaharada2004}

Several studies have demonstrated that isoelectronic sublattice alloying on the \textit{X}-site of \textit{X}NiSn (compounds in the Ti$_x$Zr$_{y}$Hf$_{1-x-y}$NiSn family) can drastically 
reduce the lattice thermal conductivity without considerably reducing electrical conductivity.\cite{hohl1999,Uher1999,sakaruda_and_shutoh2005,Shutoh2005,Kimura2009,chen2013,downie2014,galazka2014} 
There are, moreover, studies showing that introducing barriers like grain boundaries in polycrystalline samples may scatter long-wavelength acoustic phonons, thus reducing $\kappa_{\ell}$ even further,\cite{tritt2001,bhattacharya2002,Bhattacharya2008,YuZhuShiEtAl2009,Yu2009,Bhardwaj2012,xie2012_apl,downie2014}  
in particular so for nano-structured bulk samples.\cite{PbTe:nanostrucuring,Minnich2009_nanoreview,Chen2012535}

To gain insight into the role of different phonon scattering mechanisms in HH compounds, we calculate in this paper $\kappa_{\ell}$ of Ti$_x$Zr$_{y}$Hf$_{1-x-y}$NiSn compounds using density functional theory in conjunction with the Boltzmann transport equation in the relaxation time approximation.
This is combined with anharmonic scattering, calculated in the finite-displacement (frozen phonon) approach,\cite{phono3py} 
mass-disorder scattering arising from alloying of the $X$ sublattice of $X$NiSn compounds and 
the natural abundance of isotopes, as well as boundary scattering.
These calculations serve as a theoretical optimization of the alloy composition of \textit{X}NiSn,
probing the potential of combining these effects and guiding 
efforts to fabricate samples with low thermal conductivity. 
We explain the trends due to alloying in terms of a competition between the changing nature of the phonon modes and 
the mass variance. 
With an optimal alloy combination, we find that the residual heat is carried primarily by low-energetic phonons. 
Finally, we show how boundary scattering can be utilized to further reduce  $\kappa_\ell$ by efficiently scattering these phonons.

% This paper

%------------------------------------------------ %

\section{Method}\label{sec:method}

\subsection{Lattice thermal conductivity}

In the phonon Boltzmann transport 
equation (BTE) within the relaxation time approximation (RTA), \cite{physics-of-phonons}
the lattice thermal conductivity $\kappa_{\ell}$  for cubic materials is given by  
%TODO update equation.
\begin{equation}
  \kappa_{\ell} = \frac{1}{3}  \sum_{ \mathbf{q}, j} \int\frac{\mathrm{d}^3 q}{(2\pi)^3} 
C_{V,j\mathbf{q}} \tau_{j\mathbf{q}} v_{j\mathbf{q}}^2 \,.
    \label{eq:kappal}
\end{equation}
Here $j$ and $\mathbf{q}$ are the phonon band index and wavevector. %$V$ is the unit-cell volume, and $N$ the number of cells in the crystal.
The magnitude of the phonon group velocity $v_{j\mathbf{q}} $ is given by the absolute value of
the gradient of the phonon dispersion $\omega_{j\mathbf{q}}$. 
$C_{V,j\mathbf{q}} $ is the heat capacity  and $ \tau_{j\mathbf{q}} $ the relaxation time.  
In our calculations, the relaxation time includes contributions from 
anharmonic three-phonon scattering, $\tau^{\textrm{anh}}_{j\mathbf{q}}$, mass-disorder scattering $\taumd$ and grain-boundary scattering $\taubs$.  For the pure (unmixed) HH compounds $\taumd$ is made up entirely of isotope scattering. When alloying the $X$ sublattice, $\taumd$ is generally dominated 
by mass-disorder on the \textit{X} site. The total scattering rate (i.e. reciprocal relaxation time)
is given by Mathiessen's rule
\begin{equation}\label{eq:matthiesen}
  \dfrac{1}{\tau_{j\mathbf{q}}} = \dfrac{1}{\tauanh} +  \dfrac{1}{\taumd} + \dfrac{1}{\taubs}\,,
\end{equation}
where the scattering rate due to mass-disorder 
is the combined effect of isotope and alloy scattering.

\subsection{Mass-disorder scattering}\label{sec:VCA}

Isoelectric substitutions efficiently scatter the more energetic phonons, which are carrying a large part of the lattice heat.\cite{Li_PRB12}
However, a disordered solid is difficult to model from first principles. 
For instance, phonon calculations become computationally more demanding when symmetry is broken, such as in alloys, since the number of non-equivalent displacements increases dramatically. 
This makes it challenging to calculate mass-disorder scattering explicitly using atomic-scale calculations, but an effective approach is offered by the virtual crystal approximation (VCA). \cite{Abdeles1963,tamura_isotope_II,Tian2012,Garg2011}
This assumes that the electronic nature of alloys is similar to that of the 
respective pure compounds and that its properties can be calculated as an effective compound consisting of mass-averaged atoms.

We use the VCA to represent the disordered sublattice $X$ of the $X$NiSn system, with $X=$Ti$_x$Zr$_y$Hf$_{1-x-y}$, as a virtual ("effective") site $X_{\rm virtual}$. The $X_{\rm virtual}$NiSn system thus retains the symmetry of the pure crystals. 
The properties of the effective system are calculated by linearly interpolating the masses, that is $m_{\rm virtual}  = x\,m_{\rm Ti} + y\,m_{\rm Zr} + (1-x-y)\,m_{\rm Hf}$, as well as the second-, and third-order force constants of TiNiSn, ZrNiSn, and HfNiSn. 
Based on these averages we calculate the phonon dispersion and lattice thermal conductivity of the mixed compositions Ti$_x$Zr$_y$Hf$_{1-x-y}$NiSn. 

We find that the trends in the phonon dispersion and lattice thermal conductivity are not very sensitive 
to whether the forces are interpolated or if the forces of one HH compound is used to represent the virtual crystal (not shown here); the variation of the effective mass with alloy concentration is on the other hand decisive for the trends.
While still an inherently a crude approach for alloys, this supports that describing alloy scattering in terms of mass-disorder scattering within the VCA is reasonable, and will be used in the following to quantify such scattering. 

\begin{figure}[t!]
	\centering
        \includegraphics[width=0.8\linewidth]{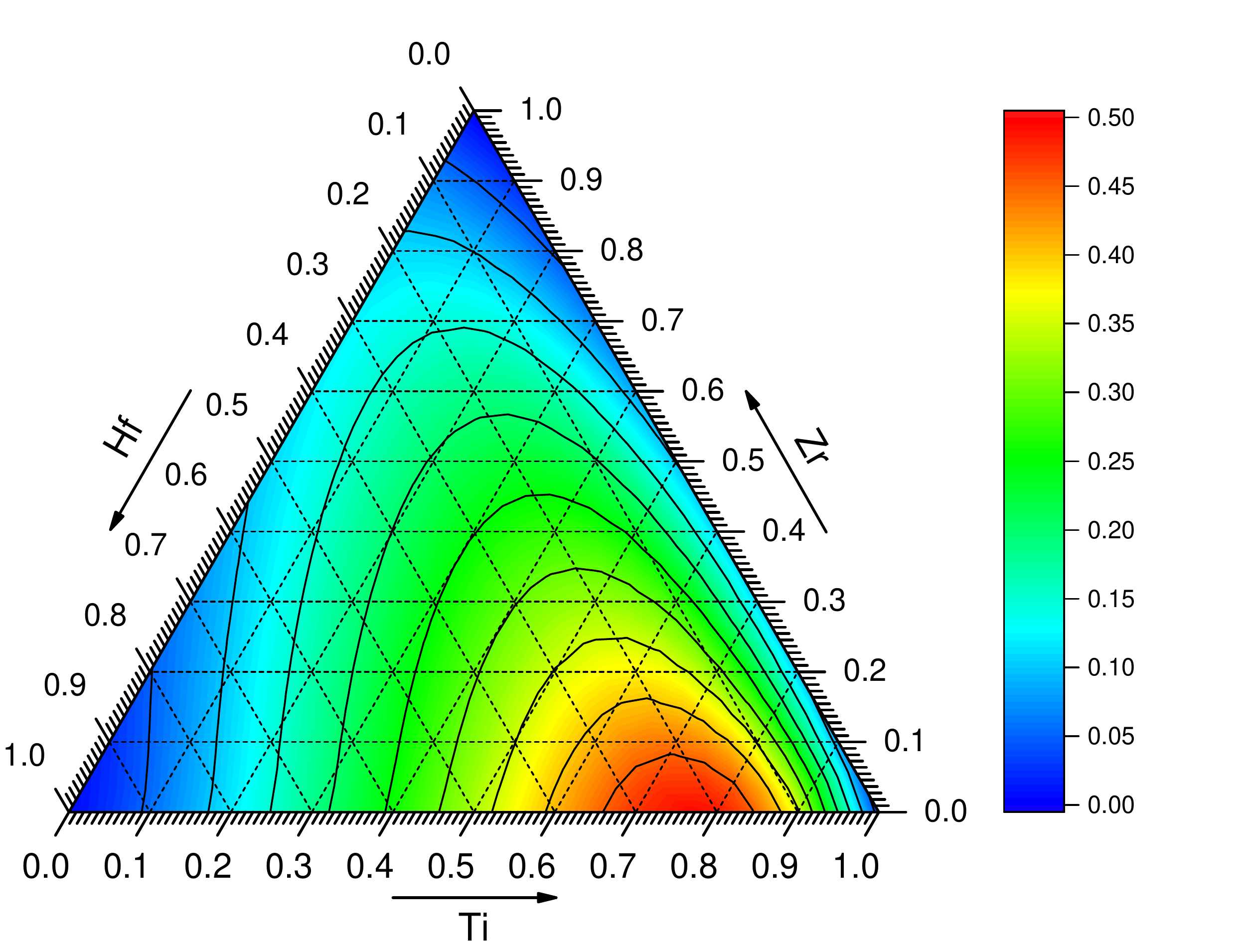}
        \caption{A color map of the $X$- site mass variance parameter $M_{\textrm{var}}$ of Ti$_x$Zr$_y$Hf$_{1-x-y}$NiSn. }
      	\label{fig:massvar}
\end{figure}

The scattering rate due to mass-disorder\cite{tamura_isotope_II}
is given by 
\begin{align}
\dfrac{1}{\tau^{\mathrm{md}}_{j\vec{q}}} =& \dfrac{\pi}{6} \omega_{j\vec{q}}^2 V  \sum_{\vec{b}} M_{\mathrm{var}}(\vec{b}) |\vec{e}(\vec{b}|j\vec{q})|^2 g_{\vec{b}}(\omega_{j\vec{q}}) \,.
\label{eq:md}
\end{align}
Here  $g_{\vec{b}}(\omega_{j\vec{q}})$ is the partial density of states of atom/site~$\vec{b}$ in the cell of the virtual crystal.
The mass variance parameter $ M_{\text{var}}(\mathbf{b}) $  is given by
\begin{equation}
	M_{\text{var}}(\mathbf{b}) = \sum_{i}f_i 
	\left(\dfrac{ \bar{m}(\mathbf{b}) - m(\mathbf{b};i)}{\bar{m}(\mathbf{b})}\right)^2,
    \label{eq:Mvar}
\end{equation}
where $f_i $ is the relative fraction of species $i$ and $ m(\mathbf{b};i) $ is the mass of atom $i$ at site ${\bf b}$.  
$ \bar{m}(\mathbf{b}) = \sum_i f_i m(\mathbf{b};i) $ is 
the average mass. $\vec{e}(\vec{b}|j\vec{q})$ is the phonon amplitude of the atom at site $\mathbf{b}$ corresponding to the phonon mode of band $j$ and wavector $\vec{q}$.

The mass variance parameter $M_{\mathrm{var}}$ corresponding to the natural distribution of isotopes\cite{Abdeles1963,tamura_isotope_II} is at most in the order of $10^{-3}$. 
It is significantly smaller than the $M_{\mathrm{var}}$ for the effective crystal on the $X$ sublattice:
Figure~\ref{fig:massvar} shows a contour map of $M_\mathrm{var}$ (eq.~\ref{eq:Mvar}) for different compositions of Ti$_x$Zr$_y$Hf$_{1-x-y}$NiSn, and it is evident that this is a significantly larger effect with values up to 0.5.
The map indicates that mass-disorder scattering due to $M_\mathrm{var}$ should be significantly stronger
along the Ti$_x$Hf$_{1-x}$NiSn line than along the two other "binary" compositional lines on the rim of the triangle. 
However, eq.~(\ref{eq:md}) shows that mass-disorder scattering is 
also affected by how $M_\mathrm{var}$ couples to the phonon mode amplitude ~$|\vec{e}(\vec{b}|j\vec{q})|$ and the partial DOS (PDOS).\cite{tamura_isotope_II}
We will return to this in Sec.\ \ref{sec:results}. 

\subsection{Anharmonic scattering}

Anharmonic  scattering is included at the
three-phonon interaction level. 
Terms beyond the third order are ignored as they are typically less 
significant at modest temperatures 
and they are costly to compute. 
The scattering rate due to
anharmonic three-phonon interactions is 
obtained by using Fermi's golden rule,\cite{ankita,phono3py}
\begin{widetext}
	\begin{equation}
	\begin{aligned}
                 \dfrac{1}{\tau_{j\mathbf{q}}^{\textrm{anh}}} &= \dfrac{36\pi}{\hbar^2} 
			\sum_{j'\mathbf{q'}j''\mathbf{q''}} |\Phi(-j\mathbf{q}, 
			j'\mathbf{q'}, j''\mathbf{q''})|^2 
			\\&\times\Bigr[(n_{j'\mathbf{q'}} + n_{j''\mathbf{q''}} + 
                          1)\delta\left(\omega_{j''\mathbf{q''}} - \omega_{j'\mathbf{q'}} - 
                          \omega_{j\mathbf{q}} \right)
			+ 2(n_{j'\mathbf{q'}} - 
                        n_{j''\mathbf{q''}})\delta\left( \omega_{j\mathbf{q}} - 
                      \omega_{j'\mathbf{q'}} - \omega_{j''\mathbf{q''}}\right)  \Bigl],
		\end{aligned}
	\end{equation}
\end{widetext}

\noindent
where  $ \Phi(-j\mathbf{q}, j'\mathbf{q'}, j''\mathbf{q''}) $ 
are the third-order force constants\cite{physics-of-phonons}
and $n_{j''\mathbf{q''}}$ is the phonon occupation of the each mode. 

\begin{figure*}[t!]
	\centering
	\includegraphics[width=1\linewidth]{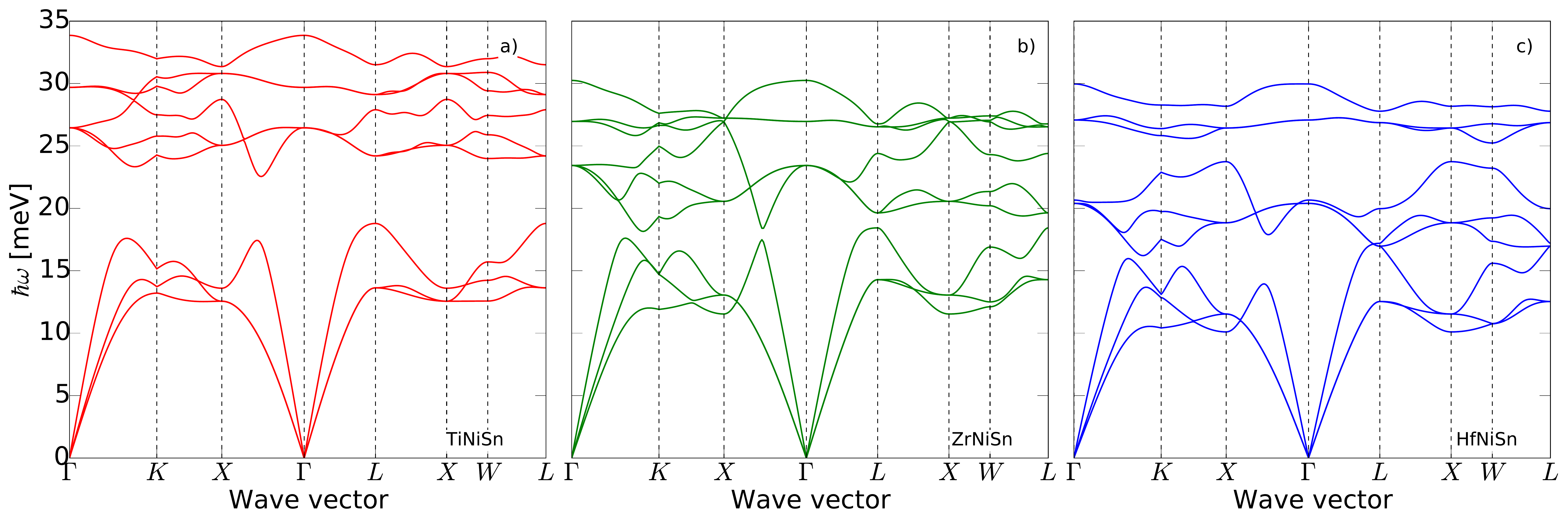}
	\caption{The phonon dispersion for a) TiNiSn, b) ZrNiSn and c) HfNiSn. 
	The curves for the three compounds are similar, but with a decreasing frequency of the three lower optical 
	phonons bands when going from TiNiSn to HfNiSn. }
	\label{fig:dispersion}
\end{figure*}

\subsection{Boundary scattering}\label{sec:boundary}

Quantifying grain boundary scattering from an atomistic point of view is demanding, given the great variety of boundaries that may exist and the difficulty of treating scattering across even one such boundary explicitly.\cite{SiGeinterface,boundary:transmisson,Kirievsky2014:gb}
Nonetheless, an estimate can be obtained with a simple model assuming purely diffusive scattering --- every phonon hitting a boundary is completely absorbed and 
re-emitted during the scattering even. \cite{ankita,madsen2014} 
The scattering rate is then given by \cite{ziman}
\begin{equation}\label{eq:bs}
  \frac{1}{\taubs} = \dfrac{v_{j\mathbf{q}}}{L}\,,
\end{equation}
where $L$ is the typical grain size, corresponding to the phonon mean free path.

\subsection{Computational details}

Electronic structure calculations used to compute second- and third-order force constants, as well as structural properties, are performed with the projector-augmented plane wave method  using the Vienna ab initio simulation package \textsc{VASP}.\cite{kresse1999} 
The exchange-correlation functional is the generalized gradient approximation (GGA) of Perdew, Burke, and Ernzerhof.\cite{pbe1996} 
To set up the %KB: keep ``the'
many atomic configurations required to compute the second- and third-order force constants within the finite-displacement approach, 
we use the software package~\textsc{PHONO3PY}.\cite{phonopy,phono3py} 
This package is subsequently used to compute phonon relaxation times and the lattice thermal conductivity.
The electronic structure calculations use a $2\times2\times2 $ supercell of the conventional cubic unit cell, representing 96 atoms. 
The plane-wave cutoff energy is $450~\mathrm{eV}$ and the 
Brillouin zone is sampled on a $3\times3\times3 $ 
Monkhorst-Pack $ \mathbf{k} $-point grid using Gaussian smearing and a smearing width of 0.05~eV.  
The interatomic forces are computed using a displacement length of $0.01\:\text{\r{A}}$.
In calculating the phonon relaxation times, we integrate over the
Brillouin zone using the tetrahedron method with the phonon modes sampled on a 
$ 30\times30\times30~ \mathbf{q} $ grid.

Convergence tests indicate 
that these numerical choices converge the phonon frequencies within 
$ \sim 0.03 $ meV and the $\kappa_{\ell}$ within $ \sim 10^{-3}$ W/mK.

%------------------------------------------------ %

\section{Results and discussion}\label{sec:results}

\subsection{Harmonic properties}\label{sec:harmonic}

\begin{figure*}[t]
	\centering
        \includegraphics[width=1\linewidth]{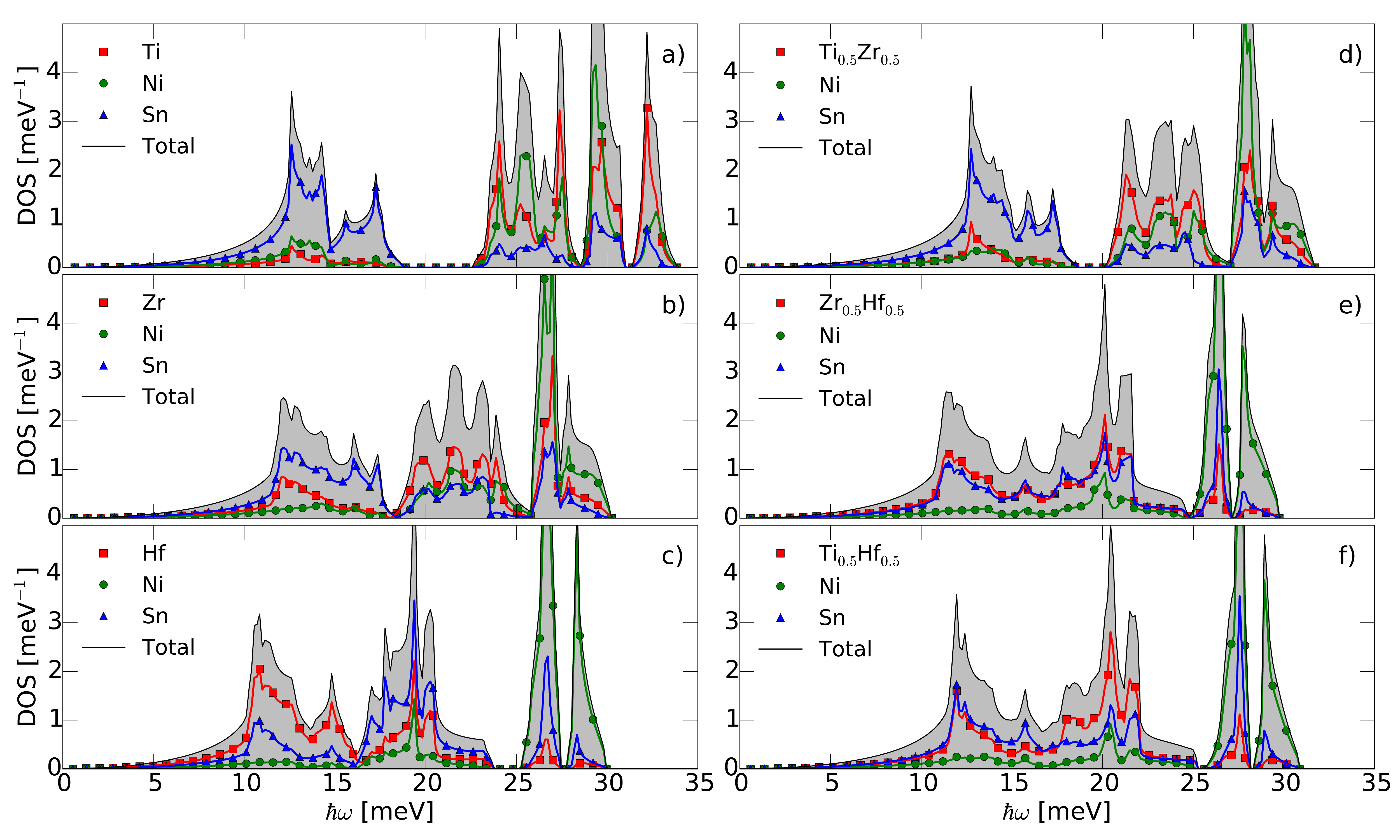}
	\caption{The full and partial phonon density of states (DOS) for a) TiNiSn, b) ZrNiSn and c) 
          HfNiSn, as well as that of the alloys d) Ti$_{0.5}$Zr$_{0.5}$NiSn, e) Ti$_{0.5}$Hf$_{0.5}$NiSn, and f)  Zr$_{0.5}$Hf$_{0.5}$NiSn obtained in the VCA. The total DOS is shown by the gray, filled area. 
        The partial DOS of the \textit{X} site (Ti, Zr, Hf, or their mixtures) is given by the red curve, whereas Ni and Sn are shown by the green and blue curves. }
	\label{fig:dos}
\end{figure*}

The phonon dispersion curves of TiNiSn, ZrNiSn, and HfNiSn are shown in Fig.~\ref{fig:dispersion}.
With three atoms per primitive cell, 
there are three acoustic and six optical phonon branches.
In ionic semiconductors such as the Half Heuslers (HHs), the longitudinal optic (LO) and transverse optical (TO) branches split
at the $\Gamma$-point. This splitting arises from the induced macroscopic polarization of these modes, an effect that is accounted for by using a non-analytic correction term in the calculations.\cite{nac_giannozzi,gonze} 

The phonon dispersion can be mapped to the phonon density of states (DOS), which is shown in Fig.~\ref{fig:dos}. 
For TiNiSn (dispersion: Fig.~\ref{fig:dispersion}a),  the acoustic and optical phonon frequencies  are separated by 
a gap throughout the Brillouin zone. 
This gap decreases with the mass at the $X$ site and is almost closed for ZrNiSn (Fig.~\ref{fig:dispersion}b), with the acoustic phonon frequencies decreased by $ <3\% $ and the lower optical phonon frequencies decreased by about $ 15\%$ compared to those of TiNiSn. 
For HfNiSn, the three lower optical phonon bands
are further decreased by about $13\%$, closing the gap completely. 
The upper three optical phonon bands are on the other hand shifted by less than $ 1\% $, thus a gap opens between the upper three and the lower three optical bands. 

Figure~\ref{fig:dos} details the partial DOS (PDOS) of the atoms in the unit cell
for each of the three pure HHs (a-c) and for the three binary 50-50 mixtures (d-f) as described in the VCA. 
The phonon modes in the 
acoustic region are dominated by vibrations of the heaviest atom in the 
\textit{X}NiSn compound; in TiNiSn and ZrNiSn this means Sn, while it is Hf in HfNiSn. 
The mass of the virtual atom $X$ in Ti$_{0.5}$Hf$_{0.5}$NiSn and Zr$_{0.5}$Hf$_{0.5}$NiSn is
113.2~u and 134.9~ u, respectively, which is similar to the Sn mass of 118.7 u.
This is reflected in similar PDOS of $X$ and Sn in those compounds, both dominating the DOS over a wide energy span ($\approx 5-22$ meV).
The optical phonon modes of highest energy are dominated 
by vibrations of the lightest atoms; hence Ti and Ni in \mbox{TiNiSn} and Ni in the other compounds. 
Correspondingly, the lowest energetic optical modes are increasingly dominated by Sn as
their frequencies are shifted downwards from TiNiSn to ZrNiSn and HfNiSn. 
As will be discussed in section~\ref{sec:alloying}, 
the changing nature of the phonon DOS impacts the strength of mass-disorder scattering.

\subsection{Thermal conductivity in bulk \textit{X}NiSn}\label{sec:bulk_MNiSn}

\begin{table}[H]
	\centering
	\caption{The calculated lattice thermal conductivity $\kappa_{\ell}$ computed in this work (boldface) compared with other theoretical calculations and experimental results in the literature.
}
	\label{tab:kappa_l}
	\begin{ruledtabular}
		\begin{tabular}{lll}
				& \multicolumn{2}{c}{$\kappa_\ell$ {[}W/mK{]}} \\
	Compound & Theory                        & Experimental        \\
  TiNiSn      & {\bf 13.8}         & 9.3 \cite{hohl1999}     \\
				& 15.4\cite{andrea}             & 7.5 \cite{Bhattacharya2008}  \\
				& 17.9\cite{carrete2014}        & 8\cite{tritt2001}    \\
				& 16.8 \cite{carrete2014}       & \\ 
				& 12.8 \cite{Hermet2016}       & \\ \\
        ZrNiSn      & {\bf 15.8}         & 8.8 \cite{hohl1999}   \\
				& 13.4 \cite{andrea}            & 10.1 \cite{muta2006} \\
				& 19.6 \cite{carrete2014}       & 6.8 \cite{Liu2015} \\
				& 17.5 \cite{carrete2014}       & 5.69 \cite{Yu2009} \\
				&                               & 4.75 \cite{Yu2009} \\ \\
        HfNiSn      &  {\bf 14.6}         & 6.7 \cite{hohl1999}  \\
				& 15.8\cite{andrea}             & 12\cite{Uher1999}   \\
				& 19.5 \cite{carrete2014}       & 6.3\cite{Liu2015}  \\
				&                               & 5.38 \cite{Yu2009}  \\
				&                               & 4.8 \cite{Yu2009}   
		\end{tabular}
	\end{ruledtabular}
\end{table}

Table~\ref{tab:kappa_l} compares the bulk calculated lattice thermal conductivities at $ 300\:\mathrm{K} $ with 
earlier published calculations and experiments.
There is a fair spread in the calculated values of $\kappa_{\ell} $,
which can partly be ascribed to different levels of theory; for instance, Andrea et al.\cite{andrea} 
solved the full BTE directly rather than using the relaxation-time approximation. 

The calculated $\kappa_{\ell}$ significantly overestimates the experimental data for the unmixed HHs (Table~\ref{tab:kappa_l}).
Recently, Katre et al.\cite{Ankita:ZrNiSn}  reported that  Ni-vacancy-antisite defect pairs have a crucial role of in reducing this difference. 
 By including Ni-vacancy-antisite scattering, obtained from density-functional theory calculations, they predicted  $\kappa_{\ell}$s
 in good agreement with experimental numbers for realistic defect concentrations. 
 For instance for ZrNiSn, with a defect-pair concentration of  2.9~\%, $\kappa_l$ is reduced to  7.4 W/m/K, and for 4~\%, it is reduced to 5.8 W/m/K.
Real samples are likely to contain significant amounts of such defects,\cite{Unraveling:TiNiSn,xie2014,Stern_NT16} 
or the related Ni interstitial defects, which may reduce the $\kappa_{\ell}$ considerably by scattering phonon modes of higher energy.\cite{Ankita:ZrNiSn}
A high solubility of Ni interstitials has been reported.\cite{TiNiSn:Ni_interst,TiNiSn:Ni_interst3,TiNiSn:Ni_interst2,ZrNiSn:Ni_interst,page}
The overestimation of  $\kappa_{\ell} $ by theory 
of the unmixed \textit{X}NiSn HHs can thus, at least partially, be explained by the presence of localized defects.
 
Another potential mechanism reducing the lattice thermal conductivity is grain boundary scattering.
In section \ref{sec:nano} we discuss how this scattering may significantly lower $\kappa_{\ell} $ when the grain size is reduced to the nanoscale. 
Metallic inclusions,  which Katayama et al.\cite{Katayama2003} reported present in \ce{TiNiSn},
could also reduce the lattice thermal conductivity.
Another possible source of discrepancy is the lack of 
fourth-order phonon-phonon scattering in the computations. Such scattering 
can lower thermal conductivity noticeably, especially at high temperatures. \cite{four-phonon}
But this does not change the fact that localized Ni-related defects appear to be the main mechanism for reducing $\kappa_{\ell}$ of the unmixed HH compounds.\cite{Ankita:ZrNiSn}

\subsection{The effect of alloying}
\label{sec:alloying}
\begin{figure}[t]
	\centering
	\includegraphics[width=1\linewidth]{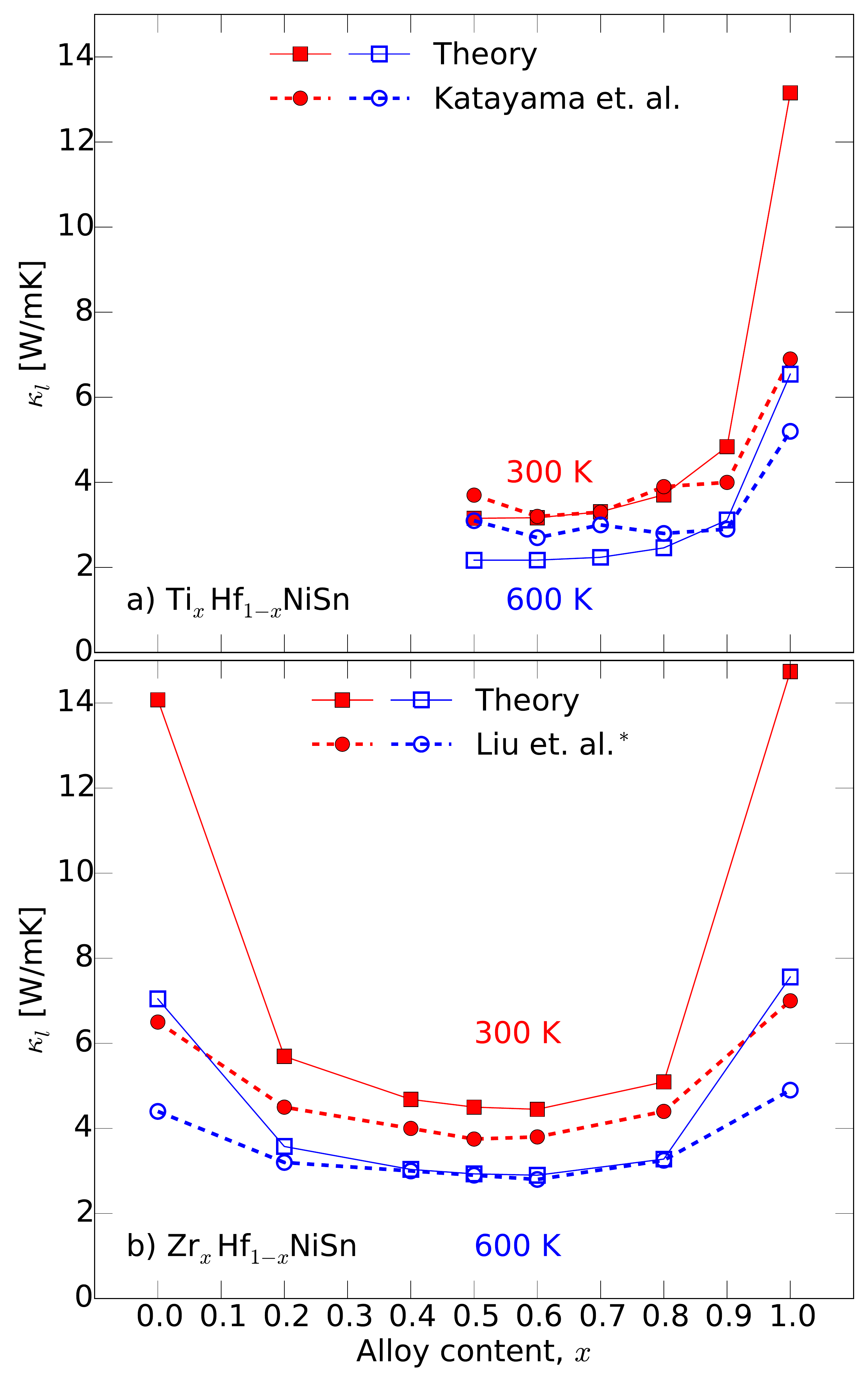}
	\caption{The lattice thermal conductivity $\kappa_{\ell}$ of the binary HH mixtures Ti$_x$Hf$_{1-x}$NiSn (a) and Zr$_x$Hf$_{1-x}$NiSn (b) as a function of composition.
    The present theoretical results are shown as squares connected with solid lines. Circles connected by dotted lines represent experimental results for 
\ce{Ti_{1-x}Hf_x NiSn} (a) by Katayama et al. \cite{Katayama2003} and \ce{Zr_{1-x}Hf_x NiSn} (b) by  Liu et al. \cite{Liu2015}  *The latter 
experiment was performed on a sample doped with 1.5\,\% Sb on the Sn site. 
The lines are a guide to the eye. The filled (unfilled) markers correspond to a temperature of $ 300\:(600)\:\mathrm{K}$. 
}
	\label{fig:binary_subplot}
\end{figure}

% Binary alloys, intro
Figure \ref{fig:binary_subplot} shows $ \kappa_{\ell} $ as a function of composition for the two binary HH alloy systems Ti$_x$Hf$_{1-x}$NiSn and Zr$_x$Hf$_{1-x}$NiSn, comparing our theoretical results with experiment. 
The result exhibits a characteristic  \textit{U}-shape of the thermal conductivity versus composition; 
that is, $ \kappa_{\ell} $ drops drastically when moving from alloy composition $x=0$ to $0.1$, whereas the change is significantly smaller in the range $ x=0.1-0.9 $.
This arises because a relatively low substitution is sufficient to scatter most of the relevant phonons.\cite{Garg2011} 
However, the drop is far less drastic for the experimental samples and 
unlike for the unmixed HHs, the computed $\kappa_{\ell}$ is in quite good 
agreement with experiment for both temperatures in Fig.\ \ref{fig:binary_subplot} (300 and $ 600\:\mathrm{K} $).
On a cautionary note, we note that the good quantitative agreement between theory and experiment might be slightly fortuitous as the VCA is an inherently crude approach as discussed in section~\ref{sec:VCA}.

To understand why the agreement between theory and experiment is so much better for the mixed compounds,
first bear in mind that the main additional scattering mechanisms 
when alloying on the $X$ sublattice is mass-disorder scattering. 
This scatters 
the more energetic phonon modes most efficiently, just like 
other localized intrinsic defects, as discussed in the previous section.\cite{Ankita:ZrNiSn}
However, the added effect of the two mechanisms is much smaller than a naive superposition of the contributions, which gives a significantly less pronounced \textit{U} shape in experiments than in the modeling without additional localized defects.

\begin{figure*}[t]
	\centering
        \includegraphics[width=1\linewidth]{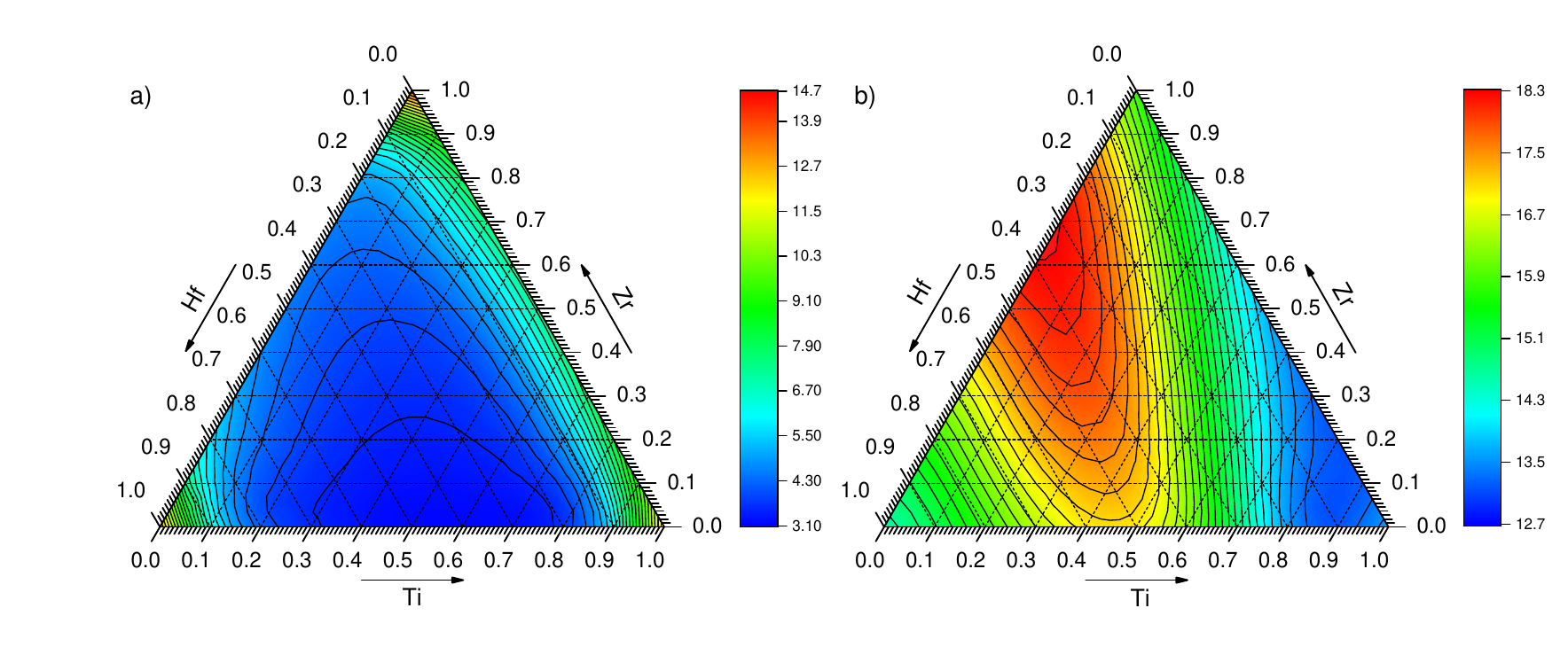}
\caption{A ternary map of $\kappa_\ell$ for the composition Ti$_x$Zr$_y$Hf$_{1-x-y}$NiSn at 300 K based on the virtual crystal approximation. The bottom right corner corresponds to TiNiSn, the top to ZrNiSn, and the bottom left to HfNiSn.
Panel a) shows the entire $ \kappa_{\ell} $, including  anharmonic phonon-phonon scattering and mass-disorder scattering.
Panel b) shows $\kappa_\ell$ when only the anharmonic phonon-phonon scattering is included through $\kappa_\ell^{\mathrm{anh}}$, and mass-disorder scattering is neglected . 
\label{fig:tern_cont}}
\end{figure*}

We proceed to calculate $\kappa_{\ell}$ for the ternary \ce{Ti_x Zr_y Hf_{1-x-y}NiSn} mixtures, where $x,y=0.1,0.2,\cdots,1.0 $ and $ x+y\leq 1 $, representing the entire ternary phase diagram.
Figure~\ref{fig:tern_cont} shows contour plots of the computed $\kappa_{\ell}$ as a function of composition. 
Panel a) presents the full $\kappa_{\ell}$ including both mass-disorder and three-phonon scattering,
whereas panel b) only includes the latter effect.
The Ti$_x$Hf$_{1-x}$NiSn compositions exhibit the lowest thermal conductivity,
with a shallow minimum around Ti$_{0.5}$Hf$_{0.5}$NiSn. 
Along the line, between  20~\%  and 80~\%~Ti, $\kappa_l$ takes values between 3.2 and 4.1 W/mK.
Moving from the optimal binary composition to ternary mixtures leads to an increase in $\kappa_l$.
Note also that the binary mixtures Ti$_{x}$Zr$_{1-x}$NiSn have a significantly higher thermal conductivity than that of corresponding Zr$_{x}$Hf$_{1-x}$NiSn compositions.

Comparing Fig.~\ref{fig:tern_cont}a) with 
the variation of the mass variance parameter $M_{\text{var}}$ (Fig.\ \ref{fig:massvar}) is a good starting point for discussing the trends in $\kappa_{\ell}$,
because of the central role played by this parameter in mass-disorder scattering; according to eq.~(\ref{eq:md}) a high $M_{\text{var}}$ should correspond to a low $\tau^{\mathrm{md}}$ and thus a small $\kappa^{\mathrm{md}}_{\ell}$.
In some ways we can see such a correlation: the highest $M_{\text{var}}$ and lowest $\kappa_{\ell}$ are found for binary Ti$_{x}$Hf$_{1-x}$NiSn, and the highest $\kappa_{\ell}$ values correspond to the pure \textit{X}NiSn compounds, where $M_{\text{var}}=0$. 
As mass-disorder scattering efficiently targets the high energetic phonons,
the more shallow minima of $\kappa_l$ along the binary mixture lines than of $M_{\text{var}}$ is thus as expected. 
It is more difficult to explain that $M_{\text{var}}$ peaks at about 80\% Ti for Ti-Hf mixing, while the minimum $\kappa_{\ell}$ is 
at 50\% Hf. Also remarkable is that $M_{\text{var}}$ is similar for Ti-Zr mixing and Zr-Hf mixing, while $\kappa_{\ell}$ is significantly higher for Ti-Zr than for Zr-Hf.
 
It is tempting to hypothesize that the remaining contribution arises from the anharmonic three-phonon scattering,
and we have plotted $\kappa_\ell^\textrm{anh}$ without mass-disorder scattering, in Fig.~\ref{fig:tern_cont}b) to test this.
It is easily concluded that this hypothesis does not hold:
In fact, $\kappa_\ell^{\rm anh}$ has a {\em maximum} around Ti$_{0.5}$Hf$_{0.5}$NiSn, precisely where we should have expected a minimal value in order to move the minimum of the full $\kappa_{\ell}$ to this point. 
% The Zr-Hf line is not at all so convincing; M_var maximum ~0.4, 
% kappa_l minimum at ~0.45, kappa^md_l maximum at ~0.35...
Similarly, 
$\kappa_\ell^{\rm anh}$ is lower along the Ti-Zr line than along Zr-Hf, which is the opposite trend of the total $\kappa_\ell$. 

To explain the trends in $\kappa_\ell^{\rm anh}$, 
we need to consider the full expression for mass-disorder scattering (eq. \ref{eq:md})\cite{tamura_isotope_II}  where the scattering strength is given by a convolution of the PDOS $g_\mathbf{b}(\omega)$, phonon-mode amplitude $e_p=|\vec{e}(\vec{b}|j\vec{q})|$, 
and mass-variance $M_{\rm var}(\vec{b})$ of a given site. 
Keep in mind that alloying in our case only increases the value of this parameter for the \textit{X}-site. 
It is then enlightening to inspect the PDOS of the six compositions shown in Fig.~\ref{fig:dos}.
% Ideally we should have moved this discussion to after showing
% where to find the main contributions, but...
We will later show that the contributions to reducing $\kappa_\ell$ originate predominantly from the acoustic modes with frequencies up to about $\sim$~16--18~meV, depending on the composition.
Let us first consider the Ti$_x$Hf$_{1-x}$NiSn mixture, where $M_{\text{var}}$ is relatively large already for a small fraction of Hf (ref.\ Fig.~\ref{fig:massvar}). 
However, for a small fraction of Hf, the $g_{X}(\omega)$ 
(and thus the amplitude $e_p$)
% can we really say that e_p follows PDOS?
in the relevant frequency range 
is modest; this can be inferred from interpolating between TiNiSn (Fig.~\ref{fig:dos}a)) and Ti$_{0.5}$Hf$_{0.5}$NiSn (Fig.~\ref{fig:dos}f)).% or in a separate plot (not shown).

We then compare the trend in the acoustic PDOS when moving from Ti to Hf along the Ti$_x$Hf$_{1-x}$NiSn line (Fig.~\ref{fig:dos}a), f), and c))
with the according trend in $M_{\rm var}$. This shows a trade-off between the maximal $M_{\rm var}$ occurring at Ti$_{0.8}$Hf$_{0.2}$NiSn and the maximal $g_X(\omega)$ (and $e_p$) occurring at HfNiSn.
This trade-off results in the minimum position of $\kappa_\ell$ being located at Ti$_{0.5}$Hf$_{0.5}$NiSn.
The PDOS of Zr$_{x}$Hf$_{1-x}$NiSn, shown in Fig.~\ref{fig:dos}e) for x=0.5, has large contributions in the entire acoustic range from the Zr-Hf virtual atom.
Thus, along the Zr-Hf line,  $\kappa_\ell$ is significantly reduced even with a relatively small $M_{\text{var}}$ (Fig.~\ref{fig:massvar}).
Finally, for Ti$_x$Zr$_{1-x}$NiSn, neither $M_{\text{var}}$ (Fig.~\ref{fig:massvar}) nor the PDOS of the Ti-Zr site is large (Fig~\ref{fig:dos}d), causing 
the  largest $\kappa_\ell$ for this binary composition.

One conclusion to draw from this analysis is that alloying should be performed on sites displaying large PDOS in the acoustic range. 
Interestingly, the crucial role of the magnitude of phonon mode nature
was noted more than three decades ago by Tamura, in the context of combining isotropically pure and impure elements in binary alloys.\cite{tamura_isotope_II} 
One promising possibility for further reduction of $\kappa_{\ell}$ in these compounds is therefore to alloy on the Sn sublattice, in order to avoid using the expensive Hf element to get a low thermal conductivity.
The PDOS of TiNiSn in Fig~\ref{fig:dos}c) indicates that the acoustic modes are dominated by vibrations of Sn, 
so introducing mass-disorder at this site might reduce the thermal conductivity considerably. Isoelectronic substitution of Sn means alloying with Si, Ge, or Pb. 
%The most beneficial would be Si because of the mass contrast as well as low price and benign environmental properties. 
%KB: This might change the mode nature too much. 
It remains to see whether the solubility of these elements is high enough to obtain a significant reduction of $\kappa_{\ell}$, and whether the other thermoelectric properties are unaffected by such substitutions.

Finally, we remark there are many reports of phase separation in Ti$_x$Zr$_y$Hf$_{1-x-y}$NiSn samples.\cite{Phase_separation,Phase_separation2,Phase_separation3}
This naturally affects the local compositions and thus 
the mean lattice thermal conductivity of phase-separated samples. 
However, the shallow minima in Fig.~\ref{fig:tern_cont}a) indicate the effect of this would be modest as long as there are stable binary or ternary mixture phases nearby in the phase diagram. 
More importantly, though, phase separation could reduce the average grain size enhancing grain boundary phonon scattering. 
In this study, we will not distinguish between different kinds of grain boundaries but rather view phase separation as one of several ways in which a polycrystalline sample could form or stabilize. As discussed in the following, small grain sizes could significantly increase grain-boundary scattering and thus lower the thermal conductivity.

\subsection{The effect of nanostructuring}\label{sec:nano}
% Intro

Mass-disorder does not efficiently scatter the low-frequency part of the acoustic phonon modes, making this range important for heat transport in mixed samples. 
These long-wavelength modes are particularly sensitive to scattering from boundaries,\cite{Goldsmid1968} 
and nanostructuring hence seems as a promising route to further reduce $ \kappa_{\ell} $ in \textit{X}NiSn alloys.\cite{Hermet2016}
Scattering from grain boundaries is in this work computed from the simple model defined by eq.~(\ref{eq:bs}). This assumes completely diffusive scattering and should give an upper bound of the scattering rate arising from boundary scattering. 

\begin{figure}[h]
  \centering
  \includegraphics[width=1.\linewidth]{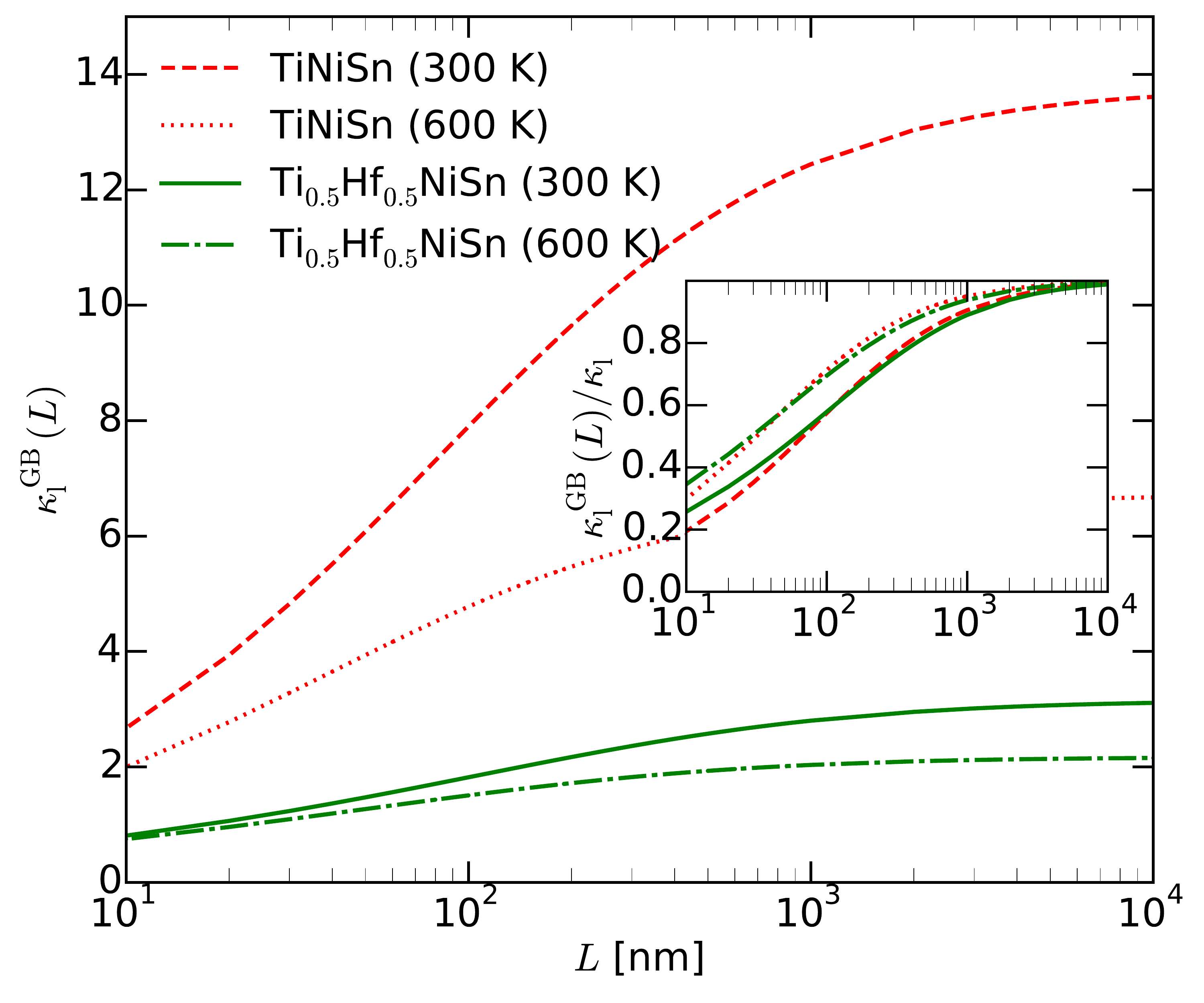}
  \caption{The total calculated thermal conductivity with grain boundary scattering included $\kappa^{\mathrm{GB}}_\ell$,
% Include ^GB in the figure
as function of average grain size $L$   for TiNiSn at temperatures 300 K and 600 K (red dashed and dotted) and Ti$_{0.5}$Hf$_{0.5}$NiSn at corresponding temperatures (green solid and dashed-dotted). 
The inset shows the lattice thermal conductivity scaled by the bulk (single crystal) value $\kappa^{\mathrm{GB}}_\ell (L)/\kappa_\ell$ for the four cases.
}
  \label{fig:kappa_vs_gs}
\end{figure}

Figure~\ref{fig:kappa_vs_gs} shows 
the lattice thermal conductivity 
at 300 and 600~K of \mbox{TiNiSn} and \mbox{Ti$_{0.5}$Hf$_{0.5}$NiSn} for different average grain sizes when grain boundary scattering is included.
$\kappa^{\mathrm{GB}}_\ell$ typically reaches the bulk (single crystal) value at a grain size of $L=10\:\mu \textrm{m}$, and is monotonously decreasing as $L$ is reduced.
The inset shows the relative reduction of $\kappa^{\mathrm{GB}}_\ell$, which is similar for both the mixed  and pure compounds. 
A grain size of 100~nm leads to a reduction of the thermal conductivity by more than 40\% at $T=300\,\mathrm{K}$ and by about a third at $T=600\,\mathrm{K}$, compared to the monocrystalline reference. A similar result was found recently by Hermet and Jund for TiNiSn.\cite{Hermet2016}
Interestingly, for grain sizes smaller than around 10 nm,  $\kappa_\ell$  becomes essentially temperature independent, particularly so for the mixed compositions. 
This can be understood from the grain-boundary scattering model (eq.\ \ref{eq:bs}) being explicitly temperature independent. For acoustic phonons in the linear regime, the model is also frequency independent. 
At higher temperatures, the heat capacity $C_{V,j,\mathbf{q}}$ in eq.~(\ref{eq:kappal}) also varies only slightly with temperature.

\begin{figure}[t!]
  \centering
  \includegraphics[width=1\linewidth]{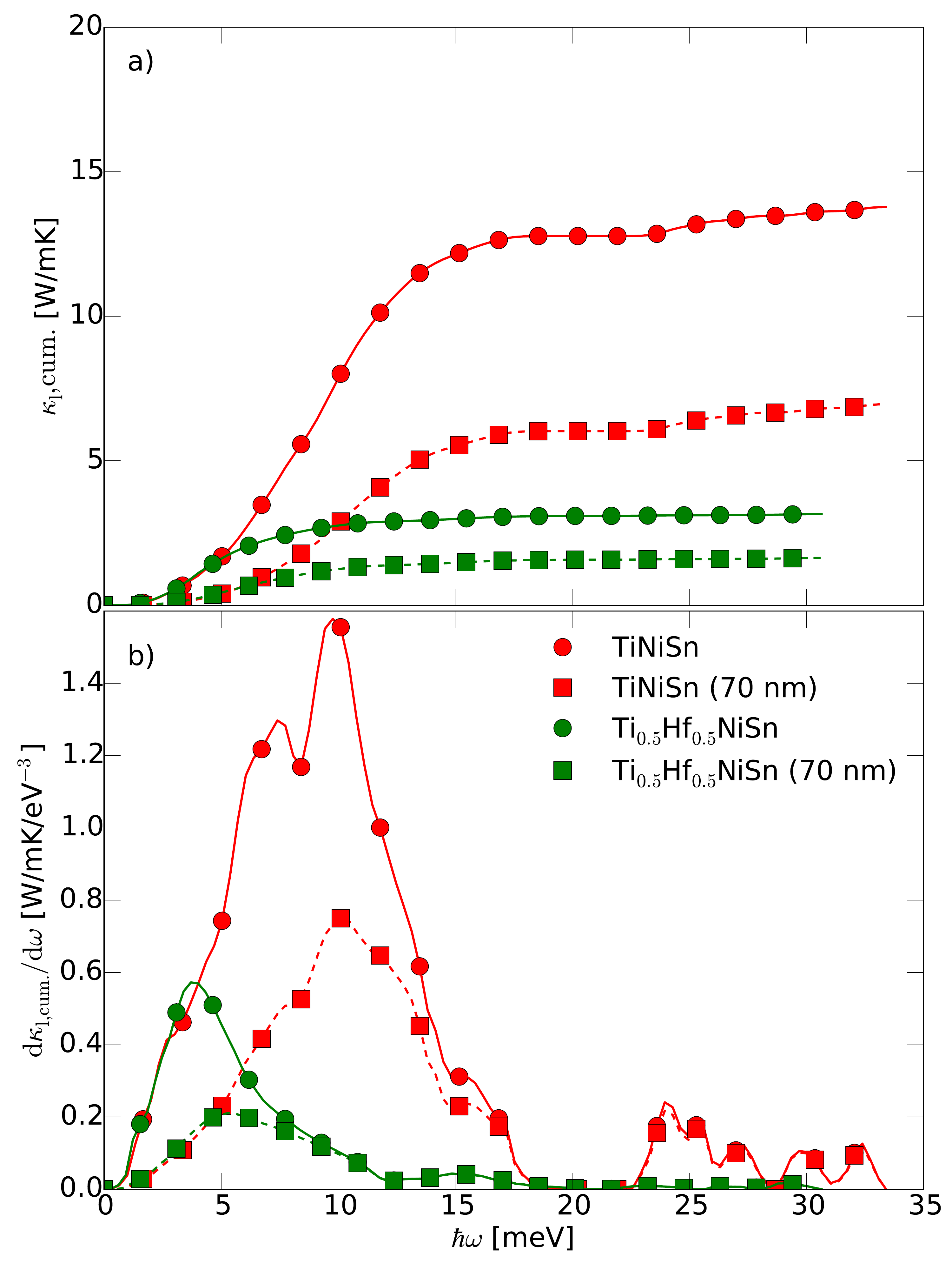}
  \caption{The cumulative $ \kappa_\ell $  at 300~K as a function of phonon frequency given in a) and its derivative in b). Pure TiNiSn with and without grain boundary scattering is shown as red squares and circles, respectively. The composition \ce{Ti_{0.5}Hf_{0.5}NiSn} with and without grain boundary scattering is shown as green squares and circles.}
  \label{fig:kaccum_alloy_boundary} 
\end{figure} 
Figure~\ref{fig:kaccum_alloy_boundary} displays 
the cumulative lattice thermal conductivity at 300~K of TiNiSn and  \ce{Ti_{0.5}Hf_{0.5}NiSn} with and without grain boundary scattering. 
It shows how, for this optimal composition, 
mass disorder scattering suppresses contributions from the most energetic phonons.
Almost all residual contributions to $ \kappa_\ell$ arise from frequencies below 10 meV, essentially the linear part of the acoustic phonon spectrum, as seen by comparing to the dispersion in Fig.~\ref{fig:dispersion}. 
An interesting feature is that even when including mass-disorder scattering, the contributions to $\kappa_\ell$ from frequencies between 3-4~meV are slightly higher in the mixed compositions.
This can be linked to the reduced anharmonic phonon scattering rates in the mixed compositions, as seen in Fig.~\ref{fig:tern_cont}.

With information about the scattering efficiency at different phonon frequencies for different scattering mechanisms, it is in principle possible to achieve maximal reduction in $\kappa_l$ with a tailored combination of mechanisms. As an example, we see in Fig.\ \ref{fig:kaccum_alloy_boundary}b) that the heat-carrying phonon modes are efficiently scattered between 5--20 meV by mass disorder on the Ti site and between 0--15 meV by grain boundary scattering. When both these mechanisms are active, the remaining contributing phonon modes are centered around 5--10 meV, and any additional engineering of phonon scattering should aim towards this region. One possibility to achieve this is to use even smaller grains sizes to scatter the acoustic region even more strongly. 
A potential drawback of smaller grains, however, is that this might also significantly reduce the electron conductivity.

\section{Conclusion}\label{sec:discussion}

This study has investigated how various scattering mechanisms lower the thermal conductivity of \textit{X}NiSn
with isoelectronic sublattice-alloying on the \textit{X}-site; $X=$(Ti, Zr, Hf). 
Good agreement was obtained between theory and experiment for mixed compositions, indicating that theory can guide efforts to optimize the thermal conductivity of these materials. 
%For unmixed samples, theory overestimates the lattice thermal conductivity, which primarily is due to localized defects. 
By screening the ternary Ti$_x$Zr$_{y}$Hf$_{1-x-y}$NiSn phase diagram, we found  
an optimum at Ti$_{0.5}$Hf$_{0.5}$NiSn with a very low thermal conductivity along the Ti$_{x}$Hf$_{1-x}$NiSn line  
with Ti content between 20 and 80\%. 
Interestingly, the overall trends, for instance, the much lower thermal conductivity along the Zr$_{x}$Hf$_{1-x}$NiSn
than the Ti$_{x}$Zr$_{1-x}$NiSn line, can only be understood from a combination of the shifting nature of the phonon modes and the magnitude of the mass variance. 

Whereas mass-disorder can be very effective at scattering high energetic phonons and even significantly impact the more energetic acoustic phonons, 
it  is less effective at scattering the acoustic phonons of lowest energy.
With a simple model of grain boundary scattering, we find that nanostructuring the sample can further reduce the thermal conductivity. 
The remaining heat is carried by acoustic phonons that are neither in the low ($<5$~meV)  nor high  ($>10$~meV) energy range. Any additional scattering mechanism should therefore target this energy range. 

Finally, our analysis demonstrates that the phonon-mode nature plays a key role in maximizing alloy scattering in these materials. This insight may pave the path to a more 
general, more deliberate strategy of optimizing the composition of alloyed 
compounds through phonon-mode engineering. Such engineering could prove crucial by moving high-througput screening of thermoelectric materials beyond the current emphasis on single unmixed compounds.

\begin{acknowledgments}
The authors gratefully acknowledge a high performance computing  allocation from the NOTUR consortium.
This work is part of the THELMA project (Project no. 228854) supported by the Research Council of Norway. This project was supported by the German Science Foundation (DFG MA 5487/2-1 and DFG MA5487/1-1).
We further thank Matthias Schrade for helpful comments.
\end{acknowledgments}

\bibliography{article.bib}

\end{document}